\documentstyle[prd,aps,psfig]{revtex}
\begin{document}
%
%
\draft
\title{Impact of $D^0 - \overline{D^0}$ mixing on\\
the experimental determination of $\gamma$}
\author{Jo\~ao P.\ Silva\footnote{Permanent address: 
	Instituto Superior de Engenharia de Lisboa,
	Rua Conselheiro Em\'{\i}dio Navarro,
	1900 Lisboa, Portugal.}
	and Abner Soffer\footnote{Permanent address: Department of Physics, 
	Colorado State University, Fort Collins, CO 80523, USA.}}
\address{Stanford Linear Accelerator Center, Stanford University, Stanford,
	 CA 94309, USA}
\date{\today}
\maketitle
\begin{abstract}
Several methods have been devised to measure the weak phase $\gamma$
using decays of the type $B^\pm \rightarrow D K^\pm$,
where it is assumed that there is no mixing in the $D^0 - \overline{D^0}$
system.
However,
when using these methods to uncover new physics,
one must entertain the real possibility that the measurements are
affected by new physics effects in the  $D^0 - \overline{D^0}$ system.
We show that even values of $x_D$ and/or $y_D$ around
$10^{-2}$ can have a significant impact in the measurement
of $\sin^2{\gamma}$.
We discuss the errors incurred in neglecting this effect,
how the effect can be checked,
and how to include it in the analysis.
\end{abstract}


\section{Introduction}
\label{sec:intro}

In the next few years,
the SM description of the charged current interactions
through the Cabibbo--Kobayashi--Maskawa (CKM) matrix \cite{CKM},
and, in particular, the nature of CP violation,
will be subject to tests of unprecedented precision.
The final objective is to over-constrain the CKM matrix and, thus,
probe any effects due to new physics.
Two important tests will be the determination of
$\sin(2 \beta)$ from the CP violating asymmetry in
$B_d \rightarrow J/\psi K_S$,
where $\beta = \arg{\left(- V_{cd} V_{cb}^\ast/ V_{td} V_{tb}^\ast\right)}$,
and the search for $B^0_s - \overline{B^0_s}$ mixing.

Interesting constraints will also arise from the decay chains
$B^\pm \rightarrow D K^\pm \rightarrow f_D K^\pm$.
The idea here is based on the fact that
\begin{equation}
\frac{A(B^+ \rightarrow D^0 K^+)}{
A(B^+ \rightarrow \overline{D^0} K^+)}
\sim
\frac{V_{ub}^\ast V_{cs}}{V_{cb}^\ast V_{us}}
\frac{a_2}{a_1}.
\label{thetrick}
\end{equation}
If the final state $f_D$ is common to $D^0$ and $\overline{D^0}$,
then these two amplitudes interfere and one probes essentially the
weak phase
$\gamma = \arg{\left(- V_{ud} V_{ub}^\ast/ V_{cd} V_{cb}^\ast\right)}$.
The phenomenological factor $|a_2/a_1| \sim 0.26$ \cite{a2a1}
accounts for the fact that the $B^+ \rightarrow D^0 K^+$ decay is color
suppressed, while the $B^+ \rightarrow \overline{D^0} K^+$ decay is not.

Inspired by a triangular construction due to Gronau and London \cite{GL},
Gronau and Wyler (GW) proposed a method to extract $\sin^2 \gamma$ 
which uses the decay chains
$B^\pm \rightarrow D K^\pm \rightarrow f_{\rm cp} K^\pm$,
where $f_{\rm cp}$ is a CP eigenstate \cite{GW}.
Atwood, Dunietz, and Soni (ADS) have modified this method,
using only final states $f_D$ which are
not CP eigenstates \cite{ADS}.
Recently,
Soffer has stressed the experimental advantages of combining
the two strategies into a single analysis,
while pointing out the complications due to the discrete ambiguities
inherent in these methods and other measurements of direct CP violation
\cite{Sof99}.

The nice feature of these decays is that they involve only
tree level diagrams and, thus, are not subject to penguin pollution.
However,
one must consider what effects the mixing in the
$D^0 - \overline{D^0}$ system might have on the
$B^\pm \rightarrow D K^\pm \rightarrow f_D K^\pm$ decay chains \cite{Mec98},
especially if these measurements are used to uncover new physics.
Otherwise,
new physics effects in the $D^0 - \overline{D^0}$ system could be
misidentified as new physics in the $B_d$ system.
In fact,
Meca and Silva \cite{Mec98} have used $x_D \sim 10^{-2}$
to argue that these effects could be of order $10\%$ in
$B^\pm \rightarrow f_D K^\pm$ decays,
and they may be as large as $100\%$ in $B^\pm \rightarrow f_D \rho^\pm$ or
$B^\pm \rightarrow f_D \pi^\pm$ decays \cite{Amo99}.

The main objective of our article is to study in detail
the effect of $D^0 - \overline{D^0}$ mixing on the
measurement of $\gamma$ in $B^\pm \rightarrow D X^\pm$ decays.
If $D^0 - \overline{D^0}$ mixing is observed, 
it must be incorporated into
the $B^\pm \rightarrow D X^\pm$ analysis. 
If only upper limits on mixing are known,
their effect should be included as a systematic error.
We also mention briefly how the effect of $D^0 - \overline{D^0}$ mixing may,
under certain conditions,
be detected in the measurement of $\gamma$.

In section \ref{sec:notation} we establish our notation.
In section \ref{sec:decays}
we present the complete expressions for the 
$B^\pm \rightarrow D X^\pm \rightarrow f_D X^\pm$ decay rates in
the presence of  $D^0 - \overline{D^0}$ mixing.
In section \ref{sec:thebigone},
we start by studying the influence of $D^0 - \overline{D^0}$ mixing
on the GW and ADS methods separately,
concentrating on some regions of parameter space.
We show that these effects depend on the specific value
of $\gamma$ and that they might affect the extraction of $\sin^2 \gamma$
by as much as $75\%$,
even for values of $x_D$ and/or $y_D$ around $10^{-2}$.
Then,
we combine the GW and ADS methods into a realistic experimental
analysis,
performing a scan over parameter space to discuss the impact that the
mixing effects have on such experiments.
We also show how to include the mixing effects in
the analysis.
We draw our conclusions in section \ref{sec:conclusions}.
For completeness,
the formulae relevant for the study of the $D^0 - \overline{D^0}$
system are included in appendix~\ref{app:A}.
Appendix~\ref{app:B} contains a comparison between CP-even
and CP-odd corrections to the extraction of $\sin^2 \gamma$.
Appendix~\ref{app:C} presents an analysis of the measurements
of the strong phases in the $D$ decays that enter in the
extraction of $\gamma$,
and which may be performed at the tau-charm factories.

\section{Assumptions and notation}
\label{sec:notation}

\subsection{Parametrizations of the decay amplitudes}

One might be surprised by the fact that
Eq.~(\ref{thetrick}),
is not invariant under the rephasing of the $u$ and $c$ quarks.
In fact,
the ratio measured experimentally in the GW and ADS methods is rather
\begin{equation}
\frac{A(B^+ \rightarrow D^0 K^+)\, A(D^0 \rightarrow f_D)}{
A(B^+ \rightarrow \overline{D^0} K^+)\, 
A(\overline{D^0} \rightarrow f_D)}.
\label{correct}
\end{equation}
This ratio depends on the weak phase in Eq.~(\ref{thetrick}),
on the relative weak phase between the decay amplitudes
$A(D^0 \rightarrow f_D)$ and $A(\overline{D^0} \rightarrow f_D)$,
and it has the correct rephasing-invariant properties.
The weak phase in Eq.~(\ref{correct}) is essentially given by $\gamma$.
Indeed,
tree level, $W$-mediated
$D$ decays only probe the weak phase in the first two families,
$\chi^\prime = \arg{\left(- V_{us} V_{ud}^\ast/ V_{cs} V_{cd}^\ast\right)}$,
and this is also the weak phase that appears with $\gamma$ in
Eq.~(\ref{thetrick}).
In the SM,
$\chi^\prime$ lies around $0.003$ radians \cite{AKL} and its presence
is completely irrelevant.
New physics could, in principle,
affect this result by altering $\chi^\prime$
or by allowing for new diagrams to drive the $D$ decays.
However,
both $\chi^\prime$ \cite{BLS} and any additional contributions to
$D$ decays \cite{Ber99} are likely to remain small in most models of
new physics.
We will neglect them henceforth.

For simplicity,
we will use the parametrizations
\begin{eqnarray}
A(B^+ \rightarrow \overline{D^0} K^+)
=
B,
& \ \ \ &
A(B^+ \rightarrow D^0 K^+)
=
\tilde{\epsilon}\, B\, e^{i\gamma} e^{i \Delta_B},
\nonumber\\ 
A(B^- \rightarrow D^0 K^-)
=
B,
& \ \ \ &
A(B^- \rightarrow \overline{D^0} K^-)
=
\tilde{\epsilon}\, B\, e^{-i\gamma} e^{i \Delta_B}\ ,
\label{param:B}
\end{eqnarray}
for the initial $B^\pm$ decays;
\begin{eqnarray}
A(\overline{D^0} \rightarrow K^+ \pi^-)
= A(D^0 \rightarrow K^- \pi^+)
&=&
A,
\nonumber\\
A(D^0 \rightarrow K^+ \pi^-)
= A(\overline{D^0} \rightarrow K^- \pi^+)
&=&
- \epsilon\, A\, e^{i \Delta_D},
\label{param:non-cp}
\end{eqnarray}
for the $D$ decays into non-CP eigenstates;
and 
\begin{equation}
A(D^0 \rightarrow f_{\rm cp})
=
A_{\rm cp},
\ \ \
A(\overline{D^0} \rightarrow f_{\rm cp})
=
\eta_f A_{\rm cp},
\label{param:cp}
\end{equation}
for $D$ decays into a CP eigenstate $f_{\rm cp}$ with CP eigenvalue
$\eta_f$.
In these parametrizations,
we have removed all irrelevant overall CP-even phases.
However,
the differences between CP-even phases in competing paths,
$\Delta_B$ and $\Delta_D$,
are physically meaningful.
$\epsilon$ and $\tilde{\epsilon}$ are discussed in the next section.

\subsection{Estimates of the parameters}

In the SM,
the value of $\gamma$ is already constrained by an overall analysis
of the unitarity triangle.
In recent reviews Ali and London find \cite{Ali99}
\begin{equation}
36^\circ \leq \gamma \leq 97^\circ
\ \ \ \Longrightarrow \ \ \ 
0.35 \leq \sin^2 \gamma \leq 1.00\ ,
\label{SMboundgamma}
\end{equation}
while Buras \cite{Bur99} quotes $44^\circ \leq \gamma \leq 97^\circ$.
The variations found in the literature are mostly due to
different estimates for the theoretical errors and to the different
methods used to combine the theoretical and experimental errors.
For definiteness,
we will use Eq.~(\ref{SMboundgamma}) as a reasonable estimate for the 
allowed region in the SM.
However,
we stress that $\gamma$ is allowed to take any value in our
analysis.
We are concerned not only with the effects that
$D^0 - \overline{D^0}$ mixing may have on SM values of $\gamma$,
but also with the possibility that such mixing effects may
`hide' new-physics by bringing values of $\gamma$ outside the
SM allowed region into this region.

Using $|V_{ub}/V_{cb}| \sim 0.08$ \cite{Ale96},
Eq.~(\ref{thetrick}),
and assuming factorization,
we get
\begin{equation}
\tilde{\epsilon} \sim |V_{ub}^\ast V_{cs}/V_{cb}^\ast V_{us}|\, 
|a_2/a_1| \sim 0.09\ .
\label{estimate:epsilonK}
\end{equation}
Notice that the parametrization in Eq.~(\ref{param:B}) is valid for any
$B^+ \rightarrow D X^+$ decay,
with $X^+ = K^+, \pi^+, \rho^+$, and higher resonances,
but the exact values of $B$, $\tilde{\epsilon}$ and $\Delta_B$
will vary from one channel to the next.
For example,
for the $B^+ \rightarrow D \pi^+$ and $B^+ \rightarrow D \rho^+$ decays
we expect
\begin{equation}
\tilde{\epsilon}_\pi \sim |V_{ub}^\ast V_{cd}/V_{cb}^\ast V_{ud}|\, 
|a_2/a_1| \sim 0.09 \times (0.22)^2 \sim 0.004 .
\label{estimate:epsilonPI}
\end{equation}
In addition,
${\cal B}[B^+ \rightarrow \overline{D^0} \rho^+]
= (1.34 \pm 0.18) \times 10^{-2}$,
${\cal B}[B^+ \rightarrow \overline{D^0} \pi^+]
= (5.3 \pm 0.5) \times 10^{-3}$ \cite{PDG},
and
${\cal B}[B^+ \rightarrow \overline{D^0} K^+]/
{\cal B}[B^+ \rightarrow \overline{D^0} \pi^+]
= 0.055 \pm 0.015$ \cite{Ath98}.
While the factor $B^2$
is roughly $18$ ($46$) times larger in
$B^+ \rightarrow D \pi^+$
($B^+ \rightarrow D \rho^+$)
decays than it is in $B^+ \rightarrow D K^+$ decays,
the decay chains of interest to us will scale instead with the
$B^2 \tilde{\epsilon}^2$ factors in the CKM-suppressed decays
\begin{eqnarray}
{\cal B}[B^+ \rightarrow D^0 K^+]
&\sim&
\tilde{\epsilon}^2 \ {\cal B}[B^+ \rightarrow \overline{D^0} K^+]
\sim 2.4 \times 10^{-6},
\nonumber\\
{\cal B}[B^+ \rightarrow D^0 \pi^+]
&\sim&
\tilde{\epsilon}_\pi^2 \ {\cal B}[B^+ \rightarrow \overline{D^0} \pi^+]
\sim 1.0 \times 10^{-7},
\nonumber\\
{\cal B}[B^+ \rightarrow D^0 \rho^+]
&\sim&
\tilde{\epsilon}_\pi^2 \ {\cal B}[B^+ \rightarrow \overline{D^0} \rho^+]
\sim 2.5 \times 10^{-7}.
\label{thescale}
\end{eqnarray}
As a result,
the $B^+ \rightarrow D K^+$ decays are the best to extract
$\sin^2 \gamma$.

Similarly,
for the decays used in the ADS method,
$\epsilon$ is the magnitude of
the ratio of the amplitudes of the doubly Cabibbo suppressed (DCS) decay 
$\overline{D^0} \rightarrow K^- \pi^+$
to the Cabibbo allowed (CA) decay $D^0 \rightarrow K^- \pi^+$.
This may be estimated from \cite{DtoKpi}
\begin{equation}
\epsilon \sim \sqrt{\frac{{\cal B}[\overline{D^0} \rightarrow K^- \pi^+]}{
{\cal B}[D^0 \rightarrow K^- \pi^+]}} \sim \sqrt{0.0031}
\sim 0.06\ .
\label{estimate:epsilon}
\end{equation}
The parametrization in Eq.~(\ref{param:non-cp}) is valid for any
$D \rightarrow f_D$ decay,
but the exact values of $A$, $\epsilon$ and $\Delta_D$
will vary from one channel to the next.
In particular,
Eq.~(\ref{param:non-cp}) reduces to Eq.~(\ref{param:cp}) 
when $\Delta_D=0$ and 
$- \epsilon = \eta_f = \pm 1$ is the CP eigenvalue of $f_{\rm cp}$.
For the CP-even eigenstates,
such as $K^+ K^-$ and $\pi^+ \pi^-$,
$\epsilon = -1$.
For the CP-odd eigenstates,
such as $K_S \pi^0$ and $K_S \phi$,
$\epsilon = +1$.

The mixing in the $D^0 - \overline{D^0}$ system may be parametrized
by the mass difference divided by the average width
($x_D \equiv \Delta m/\Gamma$),
by the width difference divided by twice the average width
($y_D \equiv \Delta \Gamma/(2 \Gamma)$),
and by a CP-violating phase ($\theta_D$) defined by
$q_D/p_D = e^{2 i \theta_D}$.
Assuming $\theta_D=0$,
\begin{equation}
x_D^2 + y_D^2 \leq \left(6.7 \times 10^{-2} \right)^2,
\label{xD-yD:limit}
\end{equation}
at the $95\%$ C.~L.\ \cite{Nel99}.
This bound is likely to remain stable even when one allows
for $\theta_D \neq 0$ \cite{Nel:private}.

Predictions for $x_D$ and $y_D$ within the SM vary considerably
among the different authors \cite{Nel:review},
but it is probably safe to estimate an upper bound around a few
times $10^{-3}$.
This uncertainty is due to the role played by SU(3) breaking effects
in the long-distance part of the calculation.
For example,
Bucella, Lusignoli and Pugliese \cite{Buc96} have estimated the SM
value for $y_D$ to lie around $1.5 \times 10^{-3}$.
However,
this result might be subject to sizeable errors,
since it comes about through a large cancellation between two
individual contributions,
each of order $3 \times 10^{-2}$.
On the other hand,
in the SM,
the value of $\theta_D$ is related to $\chi^\prime$ and is
guaranteed to be extremely small.

When one goes beyond the SM,
one finds many models for which $\theta_D$ may be large and $x_D \sim 10^{-2}$,
while $y_D$
(which is closely related to the decay rates,
where one would hardly expect any large new physics contributions)
is likely to retain its (rather uncertain) SM value
\cite{Nel:review}.
Ultimately,
these values will be determined experimentally at $B$-factories,
in fixed target experiments,
and at tau-charm factories.
We take the point of view that,
until a determination of $x_D$ and $y_D$ is available,
their upper bounds must be included as a systematic uncertainty
in the experimental determination of $\gamma$ from the
$B^\pm \rightarrow D K^\pm$ decays.

The relevant point about the notation introduced here is that
$\epsilon$ and $\tilde{\epsilon}$ are of order $10^{-1}$,
while we will take $x_D \sim y_D \sim 10^{-2}$ as 
an illustrative example.
(This is of the order of the sensitivity expected in the near future
\cite{BabarPB}.)
Therefore,
any effect proportional to $x_D/ \tilde{\epsilon}$,
$y_D/ \tilde{\epsilon}$,
$x_D/ \epsilon$,
or $y_D/ \epsilon$ is naturally of order $10\%$ \cite{Mec98},
and might be larger depending on the exact value of the other
parameters in the problem.

We should stress that the $D^0 - \overline{D^0}$ mixing effects might
be of order unity,
or even dominate,
in the $B^\pm \rightarrow D \pi^\pm$ and $B^\pm \rightarrow D \rho^\pm$
decays,
because $\tilde \epsilon_\pi \sim 0.004$ \cite{Amo99}.
As a result,
even SM values around $x_D \sim y_D \sim 10^{-3}$,
would lead to a $10\%$ effect in these channels \cite{Amo99}.

\section{The $B^\pm \rightarrow D K^\pm \rightarrow f_D K^\pm$ decay rates}
\label{sec:decays}

\begin{figure}
\centerline{\psfig{figure=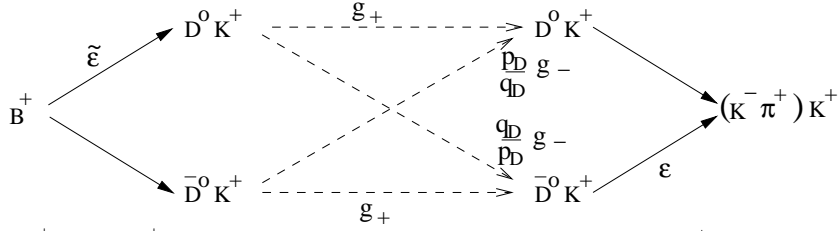,height=1.2in}}
\caption{The $B^\pm \rightarrow D K^\pm \rightarrow f_D K^\pm$
decay chain, specified for the case of $f_D = K^- \pi^+$.
For a CP eigenstate $f_D = f_{\rm cp}$ one just needs to take
$\epsilon = \pm 1$.
\label{fig:1}}
\end{figure}
The $B^+ \rightarrow D K^+ \rightarrow f_D K^+$ decay chain is
shown in Fig.~\ref{fig:1}.
The solid lines refer to
decays and the dashed lines to the time evolution of the 
$D^0 - \overline{D^0}$ system.
The functions $g_+(t)$ and $g_-(t)$ are
discussed in appendix~\ref{app:A}
and describe the flavor preserving
and flavor changing time evolutions, respectively.
The corresponding decay amplitude is obtained simply by
adding the four possible decay paths
\begin{eqnarray}
A \left(B^+ \rightarrow D K^+ \rightarrow f_D K^+ \right)
&=&
A(B^+ \rightarrow \overline{D^0} K^+)\ g_+(t)\ 
A(\overline{D^0} \rightarrow f_D)
\nonumber\\
&+&
A(B^+ \rightarrow D^0 K^+)\ g_+(t)\ A(D^0 \rightarrow f_D)
\nonumber\\*[1mm]
&+&
A(B^+ \rightarrow \overline{D^0} K^+)\ \frac{p_D}{q_D}\, g_-(t)\ 
A(D^0 \rightarrow f_D)
\nonumber\\*[1mm]
&+&
A(B^+ \rightarrow D^0 K^+)\ \frac{q_D}{p_D}\, g_-(t)\ 
A(\overline{D^0} \rightarrow f_D).
\label{master1}
\end{eqnarray}
The magnitude squared of this expression yields the time dependent decay rate.

Using the notation of section~\ref{sec:notation},
we find,
\begin{eqnarray}
& &
\frac{1}{A^2 B^2}\, 
\Gamma \left[ B^+ \rightarrow D K^+ \rightarrow f_D K^+ \right]
= 
\nonumber\\
& &
\hspace{8ex}
|g_+(t)|^2\ 
\left[ \tilde{\epsilon}^2 + \epsilon^2 - 2 \epsilon \tilde{\epsilon}
\cos ( \gamma + \Delta_B - \Delta_D) \right]
\nonumber\\
& &
\hspace{6ex}
+\ |g_-(t)|^2\ 
\left[ 1 + \epsilon^2 \tilde{\epsilon}^2 - 2 \epsilon \tilde{\epsilon}
\cos ( \gamma + 4 \theta_D + \Delta_B + \Delta_D) \right]
\nonumber\\
& &
\hspace{6ex}
+\ 2\, \mbox{Im} \left(g_+^\ast(t) g_-(t) \right)\ 
\left[ \tilde{\epsilon} ( 1 - \epsilon^2) 
\sin (\gamma + 2 \theta_D + \Delta_B)
- \epsilon ( 1 - \tilde{\epsilon}^2) 
\sin (2 \theta_D + \Delta_D)
\right]
\nonumber\\
& &
\hspace{6ex}
+\ 2\, \mbox{Re}  \left(g_+^\ast(t) g_-(t) \right)\
\left[ \tilde{\epsilon} ( 1 + \epsilon^2) 
\cos (\gamma + 2 \theta_D + \Delta_B)
- \epsilon ( 1 + \tilde{\epsilon}^2) 
\cos (2 \theta_D + \Delta_D)
\right].
\label{master}
\end{eqnarray}
To obtain the expression for the CP-conjugated decay rate,
$B^- \rightarrow D K^- \rightarrow \bar f_D K^-$,
we simply substitute $\gamma \rightarrow - \gamma$ and
$\theta_D \rightarrow - \theta_D$.
As shown in appendix~\ref{app:A},
the time integrated decay rates may be obtained through the
substitutions $|g_\pm(t)|^2 \rightarrow G_\pm$
and $g_+^\ast(t) g_-(t) \rightarrow G_{+-}$.

These expressions are completely general;
no approximations were made
(except for the use of $|q/p|=1$).
In subsequent derivations we will often simplify the expressions,
using the fact that $x_D$, $y_D$,
and $\tilde{\epsilon}$ are small when compared with one.
However the plots and estimates presented in this article were
calculated using the complete formulae.
We will also expand in $\epsilon$,
except for decays into CP-even (CP-odd) eigenstates,
where $\epsilon = - 1$ ($\epsilon = + 1$),
$\Delta_D = 0$,
and $A \rightarrow A_{\rm cp}$.

Eq.~(\ref{master}) exhibits the usual three types of CP-violating
terms.\footnote{These terms are better highlighted by 
constructing the usual CP asymmetry.
But they are also present in Eq.~(\ref{master}),
along with CP-conserving terms.}
The first line on the RHS contains a term proportional to 
$\sin{\gamma} \sin{\Delta_B}$,
which is due
to direct CP violation in the $B^\pm \rightarrow D K^\pm$ decays
(and appears multiplied by $\cos \Delta_D$ in the full decay chain).
The last line contains a term proportional to
$\sin{(2 \theta_D)} \sin{\Delta_D}$,
which is due to direct CP violation in the $D \rightarrow f_D$ decays.
In both cases,
CP violation requires a non-zero strong phase.
The usual CP violation in $D^0 - \overline{D^0}$ mixing has already been
neglected in Eq.~(\ref{master}) due to the assumption that
$|q_D/p_D|=1$.
The third line on the RHS contains a term proportional to 
$\sin{(2 \theta_D)} \cos{\Delta_D}$.
This term is due to the interference between the mixing in the 
$D^0 - \overline{D^0}$ system and its subsequent decay into the final state
$f_D$.
This term is not zero even if the strong phase $\Delta_D$ vanishes,
but it requires a nontrivial new phase in the mixing,
$\theta_D$.
We could name this a first-mix-then-decay type of interference
CP violation.

In addition,
Eq.~(\ref{master}) contains a fourth type of CP-violating
term.
This appears on the third line of Eq.~(\ref{master})
as the term involving $\sin{(\gamma + 2 \theta_D)} \cos{\Delta_B}$.
This term persists even if $\Delta_B=0$ 
(meaning that no strong phase is required),
or/and $\theta_D=0$
(meaning that no new phase is required in $D^0 - \overline{D^0}$ mixing).
In fact,
in the limit $\theta_D=\Delta_D=0$,
this term is proportional to $\sin \gamma$ and,
therefore,
it is large even within the SM.
This fourth type of CP violation was first pointed out by
Meca and Silva \cite{Mec98},
and it arises from the interference between the $B \rightarrow D$
decays and the subsequent $D^0 - \overline{D^0}$ mixing.
We might call this the first-decay-then-mix type of interference
CP violation.
Its effect is best highlighted by considering a final state
$f_D$ which tags the flavor of the $D^0$ meson
(corresponding to $\epsilon=0$ in Fig.~\ref{fig:1}).
In that case Fig.~\ref{fig:1} has only two paths
and there is still CP violation proportional to $\sin \gamma$,
meaning that it can be large even within the SM.
This example should help in clearing some frequent misconceptions
found in the literature. We see that
\begin{itemize}
\item one may in principle use a single charged $B$ decay
(and its CP conjugate) to probe a source of CP violation which
does not require strong phases,
as long as a neutral $D$ meson system is present in the intermediate state;
\item one can have a large CP-violating phase involving the
$D^0 - \overline{D^0}$ system, even within the SM.
It is true that,
in the SM,
the CP-violating phases present
in the $D^0 - \overline{D^0}$ mixing and $D$ decays are very small,
because they are proportional to $\chi^\prime$.
However,
there is a large CP-violating {\em phase} in
the first-decay-then-mix type of interference CP violation
involved in $B \rightarrow D$ decays,
even within the SM;
\item the first-decay-then-mix type of interference CP violation can,
in principle, be probed even when the $D$ meson is detected
through its (flavor-tagging) semileptonic decay.
\end{itemize}
In general these effects require values of $x_D$ and/or $y_D$ 
around $10^{-2}$,
in order to have an impact on $B^\pm \rightarrow D^0 K^\pm$ decays.
Nevertheless,
as we shall see below,
some values of the parameters are possible for which
$y_D \sim 10^{-2}$ would give a $75\%$ effect,
meaning that $y_D \sim 10^{-3}$ would still give a $7\%$ effect.

\section{The determination of $\gamma$ in the presence of
$D^0 - \overline{D^0}$ mixing}
\label{sec:thebigone}

\subsection{The no-mixing approximation}

If there were no mixing in the $D^0 - \overline{D^0}$ system,
then we would have $g_+(t) = e^{- \Gamma t/2}$
and $g_-(t) = 0$,
{\it cf.\/} appendix~\ref{app:A}.
This would leave only the uppermost and lowermost (unmixed)
decay paths of 
Fig.~\ref{fig:1};
the mixed paths represented by the diagonal dashed
time-evolution lines would be absent.
This is precisely what one assumes in both the GW and ADS methods.
In that case,
the decay amplitude reduces to the first two line of Eq.~(\ref{master1})
and the only relevant phase is that in Eq.~(\ref{correct}).
As discussed above,
the corresponding weak phase is essentially given by $\gamma$.

Obviously,
under the no-mixing approximation,
these decays are completely insensitive to any CP-violating phase
$\theta_D$ that might be present in $D^0 - \overline{D^0}$ mixing.

In the GW method,
one uses time-integrated decays rates into CP-eigenstates,
given by
\begin{eqnarray}
\Gamma \left[ B^+ \rightarrow f_{\rm cp} K^+ \right]
&\propto&
1 + \tilde{\epsilon}^2 + 2 \tilde{\epsilon}
\cos ( \gamma + \Delta_B ),
\nonumber\\
\Gamma \left[ B^- \rightarrow f_{\rm cp} K^- \right]
&\propto&
1 + \tilde{\epsilon}^2 + 2 \tilde{\epsilon}
\cos( \gamma - \Delta_B ).
\label{GWdecays}
\end{eqnarray}
Gronau and Wyler \cite{GW} assume that
$A_{\rm cp}$,
$B$,
and $\tilde{\epsilon}$
are known from the decay rates of
$D \rightarrow f_{\rm cp}$,
$B^+ \rightarrow \overline{D^0} K^+$,
and $B^+ \rightarrow D^0 K^+$,
respectively.
Therefore,
Eqs.~(\ref{GWdecays}) determine
\begin{equation}
c_+ = \cos(\gamma + \Delta_B),\ \ \mbox{and}\ \ 
c_- = \cos(\gamma - \Delta_B),
\label{c+-}
\end{equation}
from which one may extract,
\begin{eqnarray}
\sin^2 \gamma &=&
\frac{1}{2} \left[ 1 - c_+ c_- \pm \sqrt{(1-c_+^2)(1-c_-^2)} \right],
\label{trigno-exercise}
\\
\sin^2 \Delta &=&
\frac{1}{2} \left[ 1 - c_+ c_- \mp \sqrt{(1-c_+^2)(1-c_-^2)} \right].
\label{trigno-exercise-Delta}
\end{eqnarray}
As discussed below,
this determines $\gamma$ up to an eight-fold ambiguity \cite{Sof99}.

Unfortunately,
the GW method is difficult to implement for two main reasons.
The first reason is due to the hierarchy between the two
interfering amplitudes presented in Eqs.~(\ref{thetrick})
and (\ref{estimate:epsilonK}).
This is easily seen by noting that
the GW method hinges on extracting an interference of order
$\tilde{\epsilon}$ from an overall rate of order one,
as shown in Eqs.~(\ref{GWdecays}).
Since Eqs.~(\ref{GWdecays}) can be visualized as two triangles
in the complex amplitude plane,
this problem is sometimes explained by pointing out that the two
triangles are squashed.

The second reason arises from the fact that it is very
difficult to measure ${\cal B}[B^+ \rightarrow D^0 K^+]$
(and, thus, $\tilde{\epsilon}$) experimentally.
One could envision determining ${\cal B}[B^+ \rightarrow D^0 K^+]$
through the unmixed semileptonic decay chain,
$B^+ \rightarrow D^0 K^+ \rightarrow (X^- l^+ \nu_l)_D K^+$.
However,
since full reconstruction is impossible in semileptonic decays,
this process is subject to daunting combinatoric backgrounds.
Therefore,
one is led to probe ${\cal B}[B^+ \rightarrow D^0 K^+]$ through
the subsequent decay of the $D^0$ into hadronic final states,
such as $K^- \pi^+ (n \pi)^0$.
However,
as pointed out by Atwood, Dunietz and Soni \cite{ADS},
the two (unmixed) decay chains,
$B^+ \rightarrow D^0 K^+ \rightarrow (K^- \pi^+)_D K^+$
and
$B^+ \rightarrow \overline{D^0} K^+ \rightarrow (K^- \pi^+)_D K^+$
interfere at order one,
since
\begin{equation}
\frac{A(B^+ \rightarrow D^0 K^+)\, A(D^0 \rightarrow K^- \pi^+)}{
A(B^+ \rightarrow \overline{D^0} K^+)\, 
A(\overline{D^0} \rightarrow K^- \pi^+)}
\sim \frac{\tilde{\epsilon}}{\epsilon}
\sim 1.5\ .
\end{equation}
This means that it is very difficult to determine
$\tilde{\epsilon}$ and,
thus,
to implement the GW method.

Atwood, Dunietz and Soni \cite{ADS},
have turned this order one interference problem into an asset.
In their method,
one uses two final states $f_D$ for which
$\overline{D^0} \rightarrow f_D$ is a 
DCS decay,
while $D^0 \rightarrow f_D$ is Cabibbo allowed.
Examples include $f_D = K^- \pi^+$, $K^- \pi^+ \pi^0$, 
$K^- \pi^+ \pi^- \pi^+$, etc.
Then,
there are four decay rates,
\begin{eqnarray}
\Gamma \left[ B^+ \rightarrow f_{Di} K^+ \right]
&\propto&
\tilde{\epsilon}^2 + \epsilon_i^2 - 
2 \tilde{\epsilon} \epsilon_i
\cos ( \gamma + \Delta_B - \Delta_{Di}),
\nonumber\\
\Gamma \left[ B^- \rightarrow \bar f_{Di} K^- \right]
&\propto&
\tilde{\epsilon}^2 + \epsilon_i^2 - 
2 \tilde{\epsilon} \epsilon_i
\cos ( \gamma - \Delta_B + \Delta_{Di}),
\label{ADSdecays}
\end{eqnarray}
which may be solved for the four unknowns: 
$\tilde{\epsilon}$,
$\gamma$,
$\Delta_B - \Delta_{D1}$,
and $\Delta_B - \Delta_{D2}$.
For each final state,
$i=1$ or $i=2$,
one can use the analogue of Eq.~(\ref{trigno-exercise})
to obtain $\sin^2 \gamma$ and $\sin^2 (\Delta_B - \Delta_{Di})$.
These expressions, of course, depend on the unknown $\tilde{\epsilon}$,
which is determined (up to discrete ambiguities) by requiring that
the expressions for $\sin^2 \gamma$ found for $i=1$ and for $i=2$
match.
Here,
although the interference terms contribute at order
$\tilde{\epsilon} \epsilon$,
the other terms are $\tilde{\epsilon}^2$,
and $\epsilon^2$.
As a result,
the interference is of order one,
and the corresponding triangles in the complex amplitude
plane are no longer squashed.

Soffer~\cite{Sof99} has proposed to maximize the analyzing power of
the analysis by combining the GW and ADS methods,
allowing each of them to contribute the information it is most
sensitive to.
In this scheme,
one measures $\tilde{\epsilon}$, $\Delta_D$, $\Delta_B$,
and $\gamma$ by minimizing the function
\begin{equation}
\chi^2 = \sum_{m={\rm ADS, GW}} 
	\left({\Gamma^{th}_m - \Gamma^{exp}_m \over \sigma_m^{exp}}\right)^2
	+ {\rm CP\ Conj}.
	\label{eq:chi2}
\end{equation}
$\chi^2$ compares the measured integrated decay rates $\Gamma^{exp}_m$
in both the GW and ADS methods,
with their theoretical expectations,
$\Gamma_m^{th}$.
$\Gamma_m^{th}$ are functions of $\tilde{\epsilon}$,
$\Delta_B$, $\gamma$, and $\Delta_D$, 
as given by Eqs.~(\ref{GWdecays})
(for $m={\rm GW}$)
and~(\ref{ADSdecays}) (for $m={\rm ADS}$).
The comparison is done with respect to the measurement uncertainties,
$\sigma_m^{exp}$.

This analysis has several advantages over the individual GW and
ADS methods,
and, therefore,
it is most likely to be used in the actual experiment.
First,
combining the relatively high-statistics GW modes
with the small ADS decay rates improves the measurement sensitivity,
due to the addition of independent data. 
Second,
a single ADS mode is enough to measure all four unknowns,
leaving other modes to add redundancy and statistics.
Third,
this analysis is useful in removing some discrete ambiguities,
as discussed below.

\subsection{Discrete ambiguities}

We will now discuss the discrete ambiguities involved
in the determination of $\gamma$.
We start by recalling that Eq.~(\ref{trigno-exercise})
determines $\gamma$ up to an eight-fold ambiguity \cite{Sof99}.
A two-fold ambiguity arises from the fact that we know the signs
of the cosines of $\gamma + \Delta_B$ and $\gamma - \Delta_B$,
but not the signs of the corresponding sine functions.
Physically,
this amounts to a confusion between $\gamma$ and $\Delta_B$.
A further four-fold ambiguity arises from the fact that
a determination of $\sin^2 \gamma$ only determines the angle $\gamma$
up to the four fold ambiguity $\gamma$,
$\pi - \gamma$,
$\pi + \gamma$,
and $2 \pi - \gamma$.
Another way to interpret this ambiguity is to notice that
Eqs.~(\ref{GWdecays}) are invariant under the three 
independent transformations \cite{Sof99}
\begin{eqnarray}
S_{\rm sign}&:&
\ \ \ 
\gamma \rightarrow - \gamma, \ \ \ \ \ \Delta_B \rightarrow - \Delta_B;
\nonumber\\
S_{\rm exchange}&:&
\ \ \ 
\gamma \longleftrightarrow \Delta_B;
\nonumber\\
S_{\pi}&:&
\ \ \ 
\gamma \rightarrow \gamma + \pi, \ \ \ \Delta_B \rightarrow \Delta_B - \pi.
\label{discrete:ambiguities}
\end{eqnarray}
These discrete ambiguities are present in both the
GW and ADS decay rates\cite{Sof99}.
The $S_{\rm exchange}$ symmetry is $\gamma\leftrightarrow \Delta_B$
for the GW decay rates,
but $\gamma\leftrightarrow \Delta_B - \Delta_D$ for the ADS rates.
Thus,
by combining the two methods,
the $S_{\rm exchange}$ ambiguity can be resolved with a single ADS mode,
for which $\Delta_D$ is large enough.
Large values of $\Delta_D$ are expected in the SM.
By contrast,
resolving the $S_{\rm exchange}$ ambiguity in the GW method,
requires that $\Delta_B$ vary significantly from one $B$ decay mode to the
other.
This is unlikely,
because the experimental limits on CP-conserving
phases in $B \rightarrow D \pi$, $D^*\pi$, $D\rho$ and
$D^*\rho$~\cite{ref:fsi} suggest that the $\Delta_B$ are small.

The $S_{\rm sign}$ and $S_\pi$ ambiguities are likely to degrade the
value of the measurement of $\gamma$ in a non-discrete manner.
This is due to the fact that,
for $\gamma$ within the currently allowed region,
Eq.~(\ref{SMboundgamma}), 
$S_{\rm sign} S_\pi \gamma$ tends to be close enough to $\gamma$
to make the two values indistinguishable within the experimental errors.
Overall,
this results in a broad dip in the $\chi^2$ of Eq.~(\ref{eq:chi2})
as a function of the measured value of $\gamma$,
which is likely to substantially increase the measurement
uncertainty \cite{Sof99}.

It is therefore instructive to notice that the term
$\sin (\gamma + 2 \theta_D + \Delta_B)$
of Eq.~(\ref{master}) breaks the $S_{\rm sign}$ symmetry.
As a result,
one could naively expect that,
incorporating mixing into the analysis would resolve the
$S_{\rm sign}$ ambiguity and hence the $S_{\rm sign} S_\pi$ ambiguity,
bringing about significant improvement over the no-mixing case.

Unfortunately, the $S_{\rm sign}$-breaking term is too small to have a
significant effect with foreseeable data sample sizes.
This term vanishes in the large GW modes,
in which $f_D = f_{\rm cp}$,
and it accounts for about 10\% of the rate in the ADS modes
when $x_D = 0.01$.
It is estimated~\cite{Sof99} that with 600~fb$^{-1}$ collected at a
next-generation B factory,
the number of events in the ADS modes may be as large as
130\footnote{This is the case of maximal CP-asymmetry,
in which the CP-conjugate decay rate vanishes.}.
The $S_{\rm sign}$ violating term thus accounts for some 13 events,
and its contribution to the $\chi^2$ is only about $(13/\sqrt{130})^2 = 1.3$,
even when background is neglected.
Therefore,
we conclude that 60 times more data are needed to
effectively resolve the $S_{\rm sign}$ ambiguity at the $\chi^2 = 10$
level.

\subsection{The GW decay rates in the presence of mixing}

In the presence of $D^0 - \overline{D^0}$ mixing,
the $B^+ \rightarrow D K^+ \rightarrow f_{\rm cp} K^+$ decay rates 
involved in the GW method are
altered according to Eq.~(\ref{master}).
We find,
\begin{eqnarray}
\Gamma \left[ B^+ \rightarrow f_{\rm cp} K^+ \right]
&\propto&
1 + \tilde{\epsilon}^2 + 2 \eta_f\, \tilde{\epsilon}
\cos ( \gamma + \Delta_B )
\nonumber\\
&&
- \eta_f\, x_D \sin(2 \theta_D)
- y_D \left[ 2 \tilde{\epsilon} \cos{(\gamma + 2 \theta_D + \Delta_B)}
+ \eta_f\, \cos{(2 \theta_D)}
\right],
\nonumber\\
\Gamma \left[ B^- \rightarrow f_{\rm cp} K^- \right]
&\propto&
1 + \tilde{\epsilon}^2 + 2  \eta_f\, \tilde{\epsilon}
\cos ( \gamma - \Delta_B )
\nonumber\\
&&
+ \eta_f\, x_D \sin(2 \theta_D)
- y_D \left[ 2 \tilde{\epsilon} \cos{(\gamma + 2 \theta_D - \Delta_B)}
+ \eta_f\, \cos{(2 \theta_D)}
\right],
\label{GWwmix}
\end{eqnarray}
to linear order in $x_D$ and $y_D$.
Recall that $\eta_f = \pm 1$ is the CP eigenvalue of the CP
eigenstate $f_{\rm cp}$.

In these expressions we have effectively dropped the last line of
Eq.~(\ref{master1}),
because the corresponding decay chain is suppressed both by
$\tilde{\epsilon}$ and by $D^0 - \overline{D^0}$ mixing.
Moreover,
since we are only keeping linear terms in $x_D$ and $y_D$,
the second line of Eq.~(\ref{master}) does not contribute.
These approximations were made in order to simplify the expressions.
We stress that we have used the full expressions in our
analysis and in all the computer simulations.

The crucial step in the GW method is the identification
of the interference terms on the first lines of Eqs.~(\ref{GWwmix}):
$\cos(\gamma \pm \Delta_B)$.
These are identified with $c_\pm$ in Eqs.~(\ref{c+-}),
which are then used in Eq.~(\ref{trigno-exercise}).
In the GW method this interference is of order $\tilde{\epsilon}$
(although it is buried in an overall decay rate of order one),
while the leading mixing contributions are of order $x_D$ and $y_D$.
Taking $x_D \sim y_D \sim 10^{-2}$,
we expect the corrections to the determination of $\sin^2 \gamma$
to be of order $x_D/\tilde{\epsilon} \sim y_D/\tilde{\epsilon} \sim 10^{-1}$.

It is clear that the importance of the mixing terms depends
on the exact values of $\gamma$ and $\Delta_B$.
Therefore,
they could be much larger than the previous naive estimate might lead
us to believe.
Similar considerations apply to the ADS method.
As long as only upper bounds on $D^0 - \overline{D^0}$ mixing
are known,
these effects constitute systematic uncertainties that must be added
to any other experimental uncertainties.

If there is $D^0 - \overline{D^0}$ mixing,
but this is ignored in the experimental analysis,
then the $c_\pm$ extracted from the GW method become
\begin{eqnarray}
c_{+w} 
&=& 
\cos ( \gamma + \Delta_B )
- \frac{x_D}{2 \tilde{\epsilon}}
	\sin(2 \theta_D)
- \frac{y_D}{2 \tilde{\epsilon}}
	\left[ 2 \eta_f\, \tilde{\epsilon}
	\cos{(\gamma + 2 \theta_D + \Delta_B)}
	+ \cos{(2 \theta_D)}
\right],
\nonumber\\
c_{-w} &=&
\cos ( \gamma - \Delta_B )
+ \frac{x_D}{2 \tilde{\epsilon}}
	\sin(2 \theta_D)
- \frac{y_D}{2 \tilde{\epsilon}}
	\left[ 2 \tilde{\epsilon}  \eta_f\,
	\cos{(\gamma + 2 \theta_D - \Delta_B)}
	+ \cos{(2 \theta_D)}
\right],
\label{c+-GWwrong}
\end{eqnarray}
to linear order in $x_D$ and $y_D$.
The $x_D$ terms are only important if there is also a
large new CP-violating phase $\theta_D$ in the $D^0 - \overline{D^0}$ mixing.
In contrast,
the second $y_D$ term is important even if $\theta_D$ is small.
This is closely related to the fact that the linear term
in the time-dependent expression for the direct
$D \rightarrow f_{\rm cp}$ decays is proportional
to $y_D \cos{(2 \theta_D)} \pm x_D \sin(2 \theta_D)$
\cite{Ddecays}.
When Eqs.~(\ref{c+-GWwrong}) are substituted for the $c_+$ and $c_-$
in Eq.~(\ref{trigno-exercise}),
one obtains incorrect values for $\sin^2 \gamma_w$,
where $\gamma_w$ stands for the `wrong' value of $\gamma$.

Notice that the CP-odd term proportional to $x_D$ does not involve
$\Delta$.
However,
as shown in appendix~\ref{app:B},
when Eqs.~(\ref{c+-GWwrong}) are substituted into Eq.~(\ref{trigno-exercise}),
the CP-odd contributions induce an error in the determination of
$\sin^2 \gamma$ which is always proportional to 
$\sin \Delta_B$.
As a result,
the corrections due to $y_D$ tend to be more important than those
due to $x_D$, in the small $\Delta_B$ limit.

\begin{figure}
\centerline{\psfig{figure=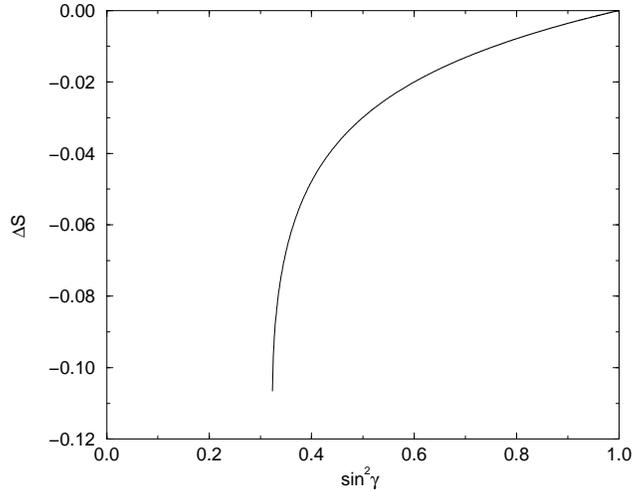,height=3in}}
\caption{Error induced by the mixing on the GW
extraction of $\sin^2 \gamma$.
Here we take 
$\tilde{\epsilon}=0.09$, $\Delta_B = 16.9^\circ$,
$\eta_f = -1$, $x_D=0.01$, $y_D=0$, and $\theta_D=30^\circ$.
\label{fig:2a}}
\end{figure}

\begin{figure}
\centerline{\psfig{figure=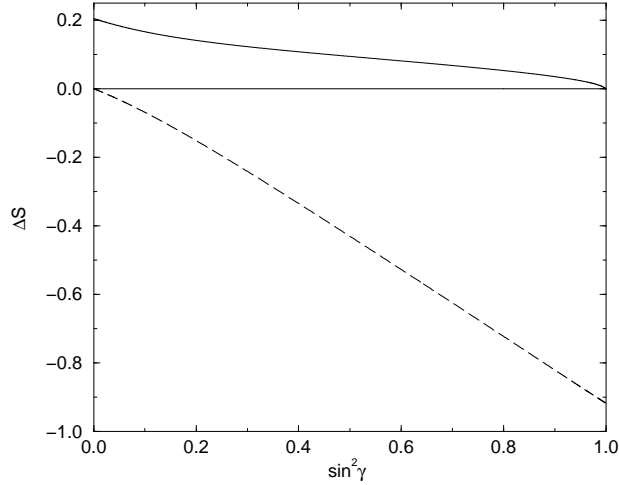,height=3in}}
\caption{Error induced by the mixing on the GW
extraction of $\sin^2 \gamma$.
Here we take 
$\tilde{\epsilon}=0.09$, $\Delta_B = 16.9^\circ$,
$\eta_f = +1$, $x_D=0$, $y_D=0.01$, and $\theta_D=0^\circ$.
\label{fig:2b}}
\end{figure}

\begin{figure}
\centerline{\psfig{figure=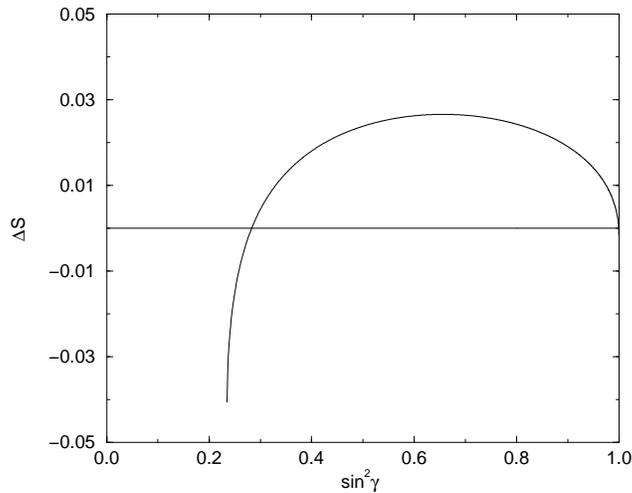,height=3in}}
\caption{Error induced by the mixing on the GW
extraction of $\sin^2 \gamma$.
Here we take 
$\tilde{\epsilon}=0.09$, $\Delta_B = 16.9^\circ$,
$\eta_f = -1$, $x_D=0.01$, $y_D=0.01$, and $\theta_D=30^\circ$.
\label{fig:2c}}
\end{figure}

These properties are illustrated in
Figs.~\ref{fig:2a},
\ref{fig:2b},
and \ref{fig:2c},
where we probe the effects due to $x_D \neq 0$, $y_D \neq 0$,
and a combination of both, respectively.
We have taken $\Delta_B = 16.9^\circ$,
which was chosen to allow a comparison with the
results in Ref.~\cite{Sof99},
and we have used the definition
\begin{equation}
\Delta S  = \sin^2 \gamma_w - \sin^2 \gamma
\end{equation}
for the difference between the `wrong' value and the `correct'
value of $\sin^2 \gamma$.

In Fig.~\ref{fig:2a}
we illustrate the effect of $x_D$ by taking
$\eta_f = -1$, $x_D=0.01$, $y_D=0$, and $\theta_D=30^\circ$.
For a true value of $\sin^2 \gamma \sim 0.35$,
we find a correction of order $-0.07$ (20\%).
Of the two possible signs in Eq.~(\ref{trigno-exercise}),
the plus sign,
corresponding to the solution plotted in the figure,
gives the correct value of $\sin^2 \gamma$,
in the limit of vanishing $x_D$.

In Fig.~\ref{fig:2b}
we illustrate the effect of $y_D$ by taking
$\eta_f = +1$, $x_D=0$, $y_D=0.01$, and $\theta_D=0^\circ$.
The two curves shown correspond to the two possible signs in
Eq.~(\ref{trigno-exercise}).
The solid (dashed) curve corresponds to the plus (minus) sign.
For each value of $\gamma$,
the line closest to the horizontal axis provides us with the
correct value for the corrections due to mixing
(the other solution is included in the discrete ambiguities).
The correct value for $\sin^2 \gamma$ is best approximated
by taking the minus sign in Eq.~(\ref{trigno-exercise})
for $\sin^2 \gamma \leq 0.19$,
while it is best approximated
by taking the plus sign in Eq.~(\ref{trigno-exercise})
for $\sin^2 \gamma \geq 0.19$.
Therefore,
the maximal value $\Delta S = \pm 0.14$ occurs precisely for 
$\sin^2 \gamma = 0.19$.
This is a correction of order 75\%, with either sign.
This example illustrates the possibility that a value of $\gamma$
outside the SM allowed range might yield a value of $\gamma_w$
inside the SM allowed region,
due to the $D^0 - \overline{D^0}$ mixing effects.

In Fig.~\ref{fig:2c}
we take 
$\eta_f = -1$, $x_D=0.01$, $y_D=0.01$, and $\theta_D=30^\circ$.
The combination of the two effects brings 
the maximum of $\Delta S$ up to higher values of
$\sin^2 \gamma$,
but the overall correction gets suppressed due to a partial cancellation
between the $x_D$ and $y_D$ terms.
For example,
for $\sin^2 \gamma \sim 0.40$,
we find a correction of order $0.02$ (5\%).
The exact form of this figure depends very sensitively on the
precise value of $\theta_D$.
As we take $\theta_D$ from $0^\circ$ to $90^\circ$,
the figures start by looking like 
Fig~\ref{fig:2b},
and end up looking like
Fig~\ref{fig:2a},
because the $x_D$ ($y_D$) effect is proportional to
$\sin 2 \theta_D$ ($\cos 2 \theta_D$).
For some values of $\theta_D \geq 90^\circ$ the $x_D$ and $y_D$ effects
come in with the same sign and there is no partial cancellation;
the figures become similar to Fig~\ref{fig:2c},
but with larger values of $\Delta S$.

Figs.~\ref{fig:2a},
\ref{fig:2b},
and \ref{fig:2c},
share some important features:
\begin{itemize}
\item The solution for $\sin^2 \gamma$ is found by using
either sign in the first of Eq.~(\ref{trigno-exercise}).
In fact,
this is the origin of the $\gamma \longleftrightarrow \Delta_B$
discrete ambiguity in the GW method.
\item The effects of $y_D$ on the GW method tend to be larger that
those due to $x_D$,
when $\Delta_B$ is very small. 
As shown in appendix~\ref{app:B},
this is the result of a $\sin \Delta_B$ suppression imposed by the
inversion procedure.
However, we should point out that this holds only under the assumption that
$\tilde{\epsilon}$ has been miraculously measured somehow.
As shown by Meca and Silva \cite{Mec98},
if one were to measure $\tilde{\epsilon}$ by tagging the
$D$ meson in the final state through its semileptonic decay,
then the $x_D$ effect would come into the extraction of $\tilde{\epsilon}$
without any $\sin \Delta_B$ suppression,
and would be as large as the $y_D$ effects.
In any case,
both effects are sizeable.
\item  The mixing effects may take values of
$\gamma$ which are outside (inside) the SM allowed region 
and yield values of $\gamma_w$ which are inside (outside) that region.
In the first case, the mixing effects hide the new physics.
In the second case,
they give a signal for new physics when, in reality,
there is none.
\item We see from Figs.~\ref{fig:2a},
\ref{fig:2b},
and \ref{fig:2c},
that,
in the presence of the mixing corrections,
not all values of $\gamma$ survive the square root used in the inversion
procedure of Eq.~(\ref{trigno-exercise}).
Indeed,
for many values of $x_D$, $y_D$, and $\theta_D$ there is a range
of values of  $\gamma$ for which $1-c_-^2<0$ or $1-c_+^2<0$.
The worst case occurs when both $1-c_-^2<0$ {\em and} $1-c_+^2<0$.
In that case,
the presence of the mixing will go undetected in the inversion
procedure, unless one is independently checking whether indeed
$ 0 \leq c_\pm^2 \leq 1$.
\item In all cases 
$\Delta S = \sin^2 \gamma_w - \sin^2 \gamma \rightarrow 0$
when $\gamma \rightarrow \pi/2$ and $\Delta_B$ is small.
This effect is explained in appendix~\ref{app:B}.
\end{itemize}

\subsection{The ADS decay rates in the presence of mixing}

Using Eq.~(\ref{master}) we obtain the
$B^+ \rightarrow D K^+ \rightarrow f_D K^+$ decay rates
relevant for the ADS method
in the presence of $D^0 - \overline{D^0}$ mixing,
as
\begin{eqnarray}
\Gamma \left[ B^+ \rightarrow f_{Di} K^+ \right]
&\propto&
\tilde{\epsilon}^2 + \epsilon_i^2 - 2 \epsilon_i \tilde{\epsilon}
\cos ( \gamma + \Delta_B - \Delta_{Di})
\nonumber\\
& &
- x_D
\left[ \tilde{\epsilon} (1 - \epsilon_i^2)
\sin (\gamma + 2 \theta_D + \Delta_B)
- \epsilon_i \sin (2 \theta_D + \Delta_{Di})
\right]
\nonumber\\
& &
-y_D
\left[ \tilde{\epsilon} \cos (\gamma + 2 \theta_D + \Delta_B)
- \epsilon_i \cos (2 \theta_D + \Delta_{Di})
\right],
\nonumber\\
\Gamma \left[ B^- \rightarrow \bar f_{Di} K^- \right]
&\propto&
\tilde{\epsilon}^2 + \epsilon_i^2 - 2 \epsilon_i \tilde{\epsilon}
\cos ( \gamma - \Delta_B + \Delta_{Di})
\nonumber\\
& &
+ x_D
\left[ \tilde{\epsilon} (1 - \epsilon_i^2)
\sin (\gamma + 2 \theta_D - \Delta_B)
- \epsilon_i \sin (2 \theta_D - \Delta_{Di})
\right]
\nonumber\\
& &
- y_D
\left[ \tilde{\epsilon} \cos (\gamma + 2 \theta_D - \Delta_B)
- \epsilon_i \cos (2 \theta_D - \Delta_{Di})
\right],
\label{ADSwmix}
\end{eqnarray}
to linear order in $x_D$ and $y_D$.
We have used here the same approximations discussed in connection
with Eqs.~(\ref{GWwmix}),
and we have also expanded in $\epsilon_i$.\footnote{Except
for the coefficient $(1 - \epsilon_i^2)$,
which was kept in order to allow a clear comparison between
Eqs.~(\ref{ADSwmix}), where $(1 - \epsilon_i^2) \sim 1$,
and Eqs.~(\ref{GWwmix}),
where $1 - \epsilon^2 = 0$ and the corresponding terms are absent.}

The crucial step in the ADS methods is the identification
of the interference terms $\cos(\gamma \pm \Delta_B \mp \Delta_D)$
on the first lines of Eqs.~(\ref{ADSwmix}).
These terms are of order $\tilde{\epsilon} \epsilon$,
while the mixing effects are of order
$x_D \tilde{\epsilon}$,
$x_D \epsilon_i$,
$y_D \tilde{\epsilon}$,
and $y_D \epsilon_i$.
As a result,
taking $x_D \sim y_D \sim 10^{-2}$,
a naive estimate predicts the mixing effects to perturb
the extraction of $\sin^2 \gamma$ at order $10\%$,
since this is the common estimate for
$x_D/\tilde{\epsilon}$,
$x_D/\epsilon_i$,
$y_D/\tilde{\epsilon}$,
and $y_D/\epsilon_i$.

If this effect is ignored in the analysis,
then the $c_\pm$ become
\begin{eqnarray}
c_{+w} 
&=& 
\cos ( \gamma + \Delta_B - \Delta_{Di})
+ \frac{x_D}{2 \epsilon_i} (1 - \epsilon_i^2)
	\sin(\gamma + 2 \theta_D + \Delta_B)
- \frac{x_D}{2 \tilde{\epsilon}}
	\sin(2 \theta_D + \Delta_{Di})
\nonumber\\
&&
+ \frac{y_D}{2 \epsilon_i}
	\cos(\gamma + 2 \theta_D + \Delta_B)
- \frac{y_D}{2 \tilde{\epsilon}}
	\cos{(2 \theta_D + \Delta_{Di})},
\nonumber\\
c_{-w} 
&=& 
\cos ( \gamma - \Delta_B + \Delta_{Di})
- \frac{x_D}{2 \epsilon_i} (1 - \epsilon_i^2)
	\sin(\gamma + 2 \theta_D - \Delta_B)
+ \frac{x_D}{2 \tilde{\epsilon}}
	\sin(2 \theta_D - \Delta_{Di})
\nonumber\\
&&
+ \frac{y_D}{2 \epsilon_i}
	\cos(\gamma + 2 \theta_D - \Delta_B)
- \frac{y_D}{2 \tilde{\epsilon}}
	\cos{(2 \theta_D - \Delta_{Di})}.
\label{c+-ADSwrong}
\end{eqnarray}
There are two new features in Eqs.~(\ref{ADSwmix}) and 
(\ref{c+-ADSwrong}),
which were not present in Eqs.~(\ref{GWwmix}) and (\ref{c+-GWwrong}).
The first is the presence of a term proportional to
$x_D \sin (\gamma + 2 \theta_D \pm \Delta_B)$.
As a result,
the $x_D$ contributions no longer require the presence
of a new CP-violating phase in the mixing $\theta_D$.
The second new feature is the presence of $\Delta_{Di}$.
These phases are expected to be large in the SM.
For some specific values of the parameters,
the magnitude, and even the sign, of $\Delta_{Di}$
will dramatically enhance the mixing effects.
In particular,
there are now effects of $x_D$ on $\sin^2 \gamma_w$
which are not proportional to $\sin \Delta_B$,
but rather to $\sin{(\Delta_B - \Delta_{Di})}$ or $\sin{\Delta_{Di}}$,
both of which may be large.
This property is discussed in detail in appendix~\ref{app:B}.
The end result is that both $x_D$ and $y_D$ have a similar
impact on the ADS method.

Strictly speaking,
in the ADS method the mixing contributions perturb the extraction of
$\sin^2 \gamma$ both directly,
as discussed above,
and indirectly,
through their effects in the determination of
$\tilde{\epsilon}$.
In order to obtain a simple estimate and to allow comparison with
the GW method,
we will also assume in this section 
that $\tilde{\epsilon} \sim 0.09$ is given
(in which case the ADS method would require only one final state).
Following Ref.~\cite{Sof99},
we use $\Delta_B = 16.9^\circ$, $\Delta_D = \pm 32.3^\circ$,
and $\epsilon \sim 0.06$.
The effects on $\Delta S$ tend to mimic those shown in 
Figs.~\ref{fig:2a},
\ref{fig:2b},
and \ref{fig:2c},
except that now the $x_D$ effects are present even when
$\theta_D = 0$,
and that all effects are typically much larger.
For example,
setting $x_D=0.01$, $y_D=0.01$, and $\theta_D=30^\circ$,
would yield $\Delta S \sim 0.1$ for the all range
$0.3 < \sin^2 \gamma < 0.8$.
Here,
there is no cancellation between the $x_D$ and $y_D$ terms.

\begin{figure}
\centerline{\psfig{figure=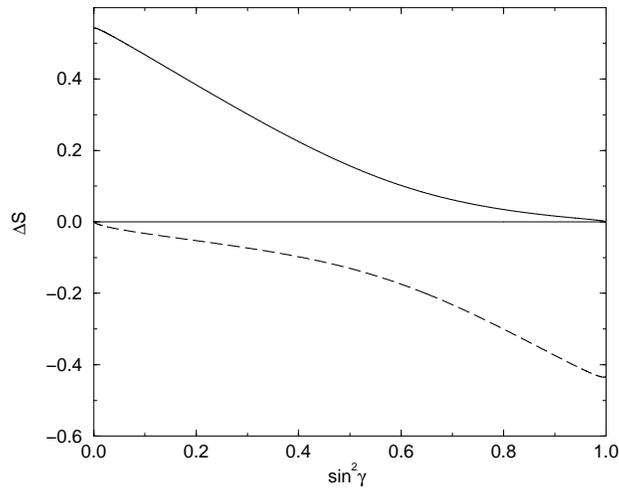,height=3in}}
\caption{Error induced by the mixing on the ADS
extraction of $\sin^2 \gamma$.
Here we take 
$\tilde{\epsilon}=0.09$, $\Delta_B = 16.9^\circ$,
$\epsilon=0.06$, $\Delta_D = - 32.3^\circ$,
$x_D=0.01$, $y_D=0$, and $\theta_D=30^\circ$.
\label{fig:3a}}
\end{figure}

\begin{figure}
\centerline{\psfig{figure=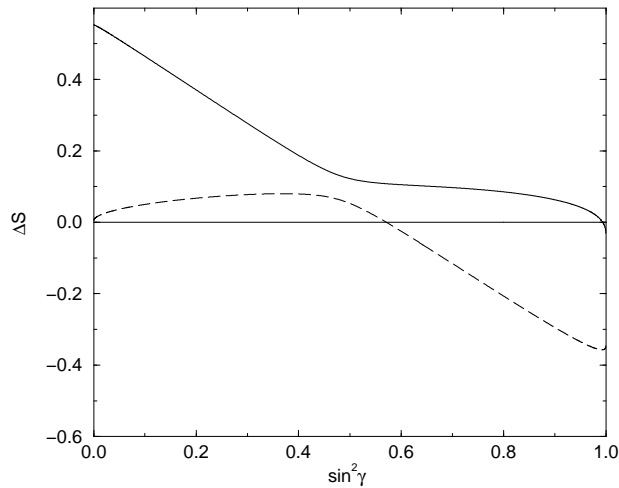,height=3in}}
\caption{Error induced by the mixing on the ADS
extraction of $\sin^2 \gamma$.
Here we take 
$\tilde{\epsilon}=0.09$, $\Delta_B = 16.9^\circ$,
$\epsilon=0.06$, $\Delta_D = - 32.3^\circ$,
$x_D=0$, $y_D=0.01$, and $\theta_D=30^\circ$.
\label{fig:3b}}
\end{figure}

More interestingly,
one can find values of the parameters which exhibit dramatic 
new features.
The solid (dashed) curves in 
Figs.~\ref{fig:3a} and \ref{fig:3b} 
correspond to the plus (minus) sign in Eq.~(\ref{trigno-exercise}).
For each value of $\gamma$,
the line closest to the horizontal axis provides us with the
correct value for the corrections due to mixing
(the other solution is included in the discrete ambiguities).
In 
Fig.~\ref{fig:3a},
we have taken $x_D=0.01$, $y_D=0$, $\theta_D=30^\circ$,
and $\Delta_D = - 32.3^\circ$.
We see that the largest correction has moved into higher
values of $\sin^2 \gamma$.
Introducing a nonzero $y_D$ would move this maximum even closer
to $\sin^2 \gamma=1$.
In 
Fig.~\ref{fig:3b},
we have taken $x_D=0$, $y_D=0.01$, $\theta_D=30^\circ$,
and $\Delta_D = - 32.3^\circ$.
Now the corrections are large and positive
for $\sin^2 \gamma < 0.57$,
they are negative for $0.57 < \sin^2 \gamma < 0.68$,
and they are positive again for  $\sin^2 \gamma > 0.68$.
This peculiar effect is due to the fact that
$\Delta S$ is minimized in different regions of $\sin^2 \gamma$
by each of the two possible signs for $\sin^2 \gamma_w$
in Eq.(\ref{trigno-exercise}).
The maximum deviation is $\pm 0.1$ and occurs for 
$\sin^2 \gamma = 0.68$.
This maximum occurs for higher values of $\sin^2 \gamma$
and, therefore,
it is proportionally smaller (15\%).
We can increase this maximum by changing $x_D$ to $0.01$.
In that case, the first $\Delta S > 0$ portion of 
Fig.~\ref{fig:3b}
is very suppressed, and it turns negative at $\sin^2 \gamma \sim 0.22$.
The maximum deviation would occur at $\sin^2 \gamma \sim 0.63$,
but would be enhanced to $\pm 0.15$ (a $24\%$ effect).

We have not shown ADS figures with large effects (in percentage)
because they do not show new features.
Nevertheless,
such possibilities do exist.
For example,
if we take $\Delta_D = + 32.3^\circ$,
$x_D=0.01$, $y_D=0.01$, and $\theta_D=60^\circ$,
we obtain a figure which is very similar to 
Fig.~\ref{fig:2b},
except that large deviations around $0.1$ extend over a much larger
domain,
going approximately from $\sin^2 \gamma \sim 0.1$ up to
$\sin^2 \gamma \sim 0.6$.
In this case,
the largest deviation occurs at $\sin^2 \gamma = 0.18$
and is $\pm 0.13$ (a $70\%$ effect).

\subsection{Combining the ADS and GW methods}
\label{sec:right}

To most closely simulate the actual experiment,
we will now combine the ADS and GW methods,
using Eq.~(\ref{eq:chi2}).
We have seen above that the GW and ADS methods are affected
differently by non-trivial mixing.
One may therefore expect that when the two methods are combined,
the total effect will be smaller than in the worst-case scenario
for either method.

To calculate $\gamma_w - \gamma$ in this scheme,
we vary $\Delta_D$, $\Delta_B$, $\gamma$, $\theta_D$,
$x_D$ and $y_D$ over their allowed ranges.
For each set of input values,
we calculate $\Gamma_m^{exp}$ of Eq.~(\ref{eq:chi2}) using the
correct expression, Eq.~(\ref{master}),
but conduct the `wrong' analysis by using Eqs.~(\ref{GWdecays})
and~(\ref{ADSdecays}) to calculate $\Gamma_{\rm GW}^{th}$ and
$\Gamma_{\rm ADS}^{th}$, 
respectively.
Having thus neglected mixing in our analysis,
we proceed to minimize Eq.~(\ref{eq:chi2}) to obtain a
measurement of $\gamma_w$.\footnote{As in \cite{Sof99},
the input value of $\tilde{\epsilon}$ is taken from 
Eq.~(\ref{estimate:epsilonK}),
but its `experimental' output value is then found
(together with the other observables)
when we minimize Eq.~(\ref{eq:chi2}).}

\begin{figure}
\centerline{\psfig{figure=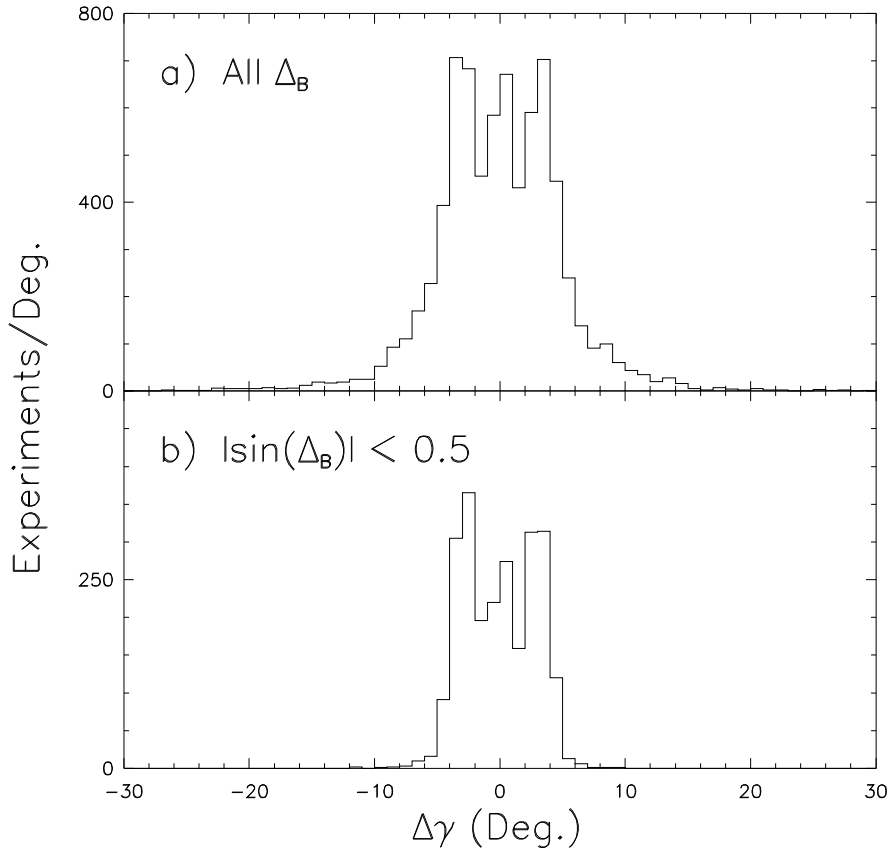,height=4.5in}}
\caption{Distributions of $\Delta\gamma = \gamma_w - \gamma$,
	the error in the measurement of $\gamma$ due to failure to 
	incorporate mixing into the analysis. Parameters are varied in 
	the ranges 
	$-180^\circ < \Delta_D < 180^\circ$,
	$36^\circ < \gamma < 97^\circ$,
	$\theta_D = 0^\circ,\, 30^\circ,\, \mbox{or } 45^\circ$
	$x_D = 0\ \mbox{or } 0.01$, and
	$y_D = 0\ \mbox{or } 0.01$.
	$\Delta_B$ is varied within
	$-180^\circ < \Delta_D < 180^\circ$ (a) or
	$|\sin(\Delta_B)|<0.5$ (b).
	}
\label{Delta-gamma}
\end{figure}

The resulting distributions of $\Delta\gamma = \gamma_w - \gamma$
are shown in Fig.~\ref{Delta-gamma}.
In Fig.~\ref{Delta-gamma}b we restrict $\Delta_B$ to the more
likely range $|\sin(\Delta_B)|<0.5$.
Since the actual values of the phases are not known,
it is not meaningful to analyze the distributions of
Fig.~\ref{Delta-gamma} in detail.
These distributions clearly indicate,
however,
that neglecting to account for $D^0-\overline{D^0}$ mixing may
significantly bias the measurement of $\gamma$.

As we have seen before,
the measurement of $\gamma$ depends on the
the precise values of the parameters in the $D$ system.
Given the current bounds on $x_D$ and $y_D$,
we expect the measurements of $\epsilon$ and $\Delta_D$ to be rather
insensitive to $D^0 - \overline{D^0}$ mixing.
The measured value of $\epsilon$,
{\it c.f.\/} Eq.~(\ref{estimate:epsilon}),
is already used in the ADS method.
Once $\Delta_D$ is measured,
{\it c.f.\/} appendix~\ref{app:C},
it can also be used as a known input parameter
in the fit for $\gamma$.
We have stressed the fact that these decays also depend on
$x_D$, $y_D$, and $\theta_D$.
The more we know about these quantities,
the better will be our bound on $\gamma$ from the
$B^\pm \rightarrow D K^\pm$ decays.

\section{Conclusions}
\label{sec:conclusions}

The decay chains $B^\pm \rightarrow D K^\pm \rightarrow f_D K^\pm$
provide a good opportunity to determine the CKM phase $\gamma$.
Naturally,
information about the $D$ decays must be included in the analysis,
either as parameters to be determined from the overall fit or as 
fixed quantities known from other $D$ system experiments.
This is well appreciated for the DCS decay parameter $\epsilon$ and
for the strong phase $\Delta_D$.

In this article,
we stress the fact that this is also true for the 
parameters involved in $D^0 - \overline{D^0}$ mixing.
This point cannot be (as it often is) overlooked,
because the extraction of $\gamma$ hinges on measurements of
small quantities.
In the GW method there is a small interference;
in the ADS method the decay rates are small.
As a result,
mild values of $x_D$ and/or $y_D$ can have an important effect
in these methods.

We have shown that dramatic effects are indeed present when the
GW and ADS methods are used individually.
In general,
combining these methods reduces the errors involved,
but one does still find deviations of order
$\gamma_w - \gamma \sim 10^\circ$.
These effects may simulate the presence of new physics
by taking values of $\gamma$ inside the SM allowed region
into values of $\gamma_w$ outside that region.
They may also obscure the presence of new physics
by taking new physics values of $\gamma$ outside the
SM allowed region and yielding $\gamma_w$ inside that region.
The importance of this error in the determination of $\gamma$
is made more acute by the discrete ambiguities associated with the
GW and ADS methods.

As a result
\begin{itemize}
\item the determination of $\gamma$ with the decays 
$B \rightarrow D K$ must be made in connection with 
the search for mixing in the $D^0 - \overline{D^0}$
mixing;
\item any uncertainties due to lack of knowledge of
$x_D$, $y_D$, and $\theta$
must be correctly included as systematic uncertainties
to the extraction of $\sin^2 \gamma$.
\item once $x_D$, $y_D$, and $\theta$ are determined 
experimentally,
they can be included in the analysis as known parameters.
\end{itemize}

\acknowledgments

We are indebted to H.\ N.\ Nelson
for elucidating remarks on the CLEO measurements of $x_D$ and $y_D$.
It is a pleasure to thank Y.\ Grossman and H.\ R.\ Quinn for
several discussions, suggestions, and for reading this manuscript.
The work of J.\ P.\ S.\ is supported in part by Fulbright,
Instituto Cam\~oes, and by the Portuguese FCT, under grant
PRAXIS XXI/BPD/20129/99	and contract CERN/S/FIS/1214/98.
The work of A.\ S.\ is supported by Department of Energy contracts
DE-AC03-76SF00515 and DE-FG03-93ER40788.


\appendix

%

\section{Time dependent evolution of the $D^0 - \overline{D^0}$ 
system.}
\label{app:A}

Assuming CPT invariance,
the mass eigenstates of the $D^0 - \overline{D^0}$ system
are related to the flavor eigenstates by
\begin{eqnarray}
| D_H \rangle &=& p_D | D^0 \rangle + q_D | \overline{D^0} \rangle\ ,
\nonumber\\
| D_L \rangle &=& p_D | D^0 \rangle - q_D | \overline{D^0} \rangle\ ,
\label{eigenvectors}
\end{eqnarray}
with $|p_D|^2+|q_D|^2=1$ and
\begin{equation}
\frac{q_D}{p_D}
= \frac{\Delta m -\frac{i}{2} \Delta\Gamma}{2 R_{12}}=
\sqrt{\frac{R_{21}}{R_{12}}}\ ,
\end{equation}
where $\Delta m = m_H - m_L$ ($H$-heavy, $L$-light) is positive
by definition,
$\Delta\Gamma = \Gamma_H - \Gamma_L$,
and  $R_{12}$ is the off-diagonal matrix element in the effective
time evolution in the  $D^0 - \overline{D^0}$ space.

Consider a $D^0$ 
($\overline{D^0}$)
meson which is created at time $t=0$ and denote by $D^0(t)$
($\overline{D^0}(t)$)
the state that it evolves into after a time $t$,
measured in the rest frame of the meson $D$.
One finds
\begin{eqnarray}
| D^0(t) \rangle &=& g_+(t) | D^0 \rangle +
\frac{q_D}{p_D} g_-(t) | \overline{D^0} \rangle,
\nonumber\\*[3mm]
| \overline{D^0}(t) \rangle &=& \frac{p_D}{q_D} g_-(t) | D^0 \rangle
+ g_+(t) | \overline{D^0} \rangle,
\label{timeevolution}
\end{eqnarray}
where
\begin{equation}
g_{\pm}(t) \equiv \frac{1}{2}
\left(
e^{-i \mu_H t} \pm e^{-i \mu_L t}
\right),
\end{equation}
$\mu_H \equiv m_H - i \Gamma_H/2$,
and $\mu_L \equiv m_L - i \Gamma_L/2$.

It is useful to trade the mass and width differences for the
dimensionless parameters $x \equiv \Delta m/\Gamma$ and
$y \equiv \Delta \Gamma/(2 \Gamma)$,
where $\Gamma = (\Gamma_H + \Gamma_L)/2$ is the average width.
We already know from studies of the direct $D$ decays,
$D \rightarrow K^+ \pi^-$, that
$x_D^2 + y_D^2 \leq \left(6.7 \times 10^{-2} \right)^2$,
at the $95\%$ C.~L.\ \cite{Nel99}.
Therefore,
we may expand the time-dependent functions as
\begin{eqnarray}
g_+(t)
&\sim&
e^{-imt}\, e^{-\tau/2}
\left[ 1 + (x_D - i y_D)^2 \tau^2/4 + \cdots \right],
\nonumber\\
g_-(t)
&\sim&
e^{-imt}\, e^{-\tau/2}
\left[ (- i x_D - y_D) \tau/2 + \cdots \right],
\end{eqnarray}
where $\tau = \Gamma t$ is the (proper) time of the $D$
evolution, in units of the $D$ average width.

The time-dependent decay rates involve
\begin{eqnarray}
|g_+(t)|^2 &=& 
\frac{e^{- \Gamma t}}{2}
\left[
\cosh \frac{\Delta \Gamma t}{2} + \cos \left( \Delta m t \right)
\right]
\sim
e^{- \tau}
\left[
1 + (y_D^2 - x_D^2) \tau^2/4 + \cdots
\right],
\label{gpquad2}
\\*[3mm]
|g_-(t)|^2 & = &
\frac{e^{- \Gamma t}}{2}
\left[
\cosh \frac{\Delta \Gamma t}{2} - \cos \left( \Delta m t \right)
\right]
\sim
e^{- \tau}
\left[
(y_D^2 + x_D^2) \tau^2/4 + \cdots
\right],
\label{gmquad2}
\\*[3mm]
g_+^\ast(t) g_-(t) & = &
\frac{1}{4} \left[
e^{- \Gamma_H t} - e^{- \Gamma_L t} - 2 i e^{- \Gamma t} \sin(\Delta m\, t)
\right]
\sim
e^{- \tau}
\left[
(- y_D - i x_D) \tau/2 + \cdots
\right].
\label{g+*g-}
\\*[3mm]
\end{eqnarray}
The time integrated decay rates involve
\begin{eqnarray}
G_+ &\equiv& \int_0^{+ \infty} \!|g_+(t)|^2 dt =
\frac{1}{2 \Gamma} \left( \frac{1}{1 - y^2} + \frac{1}{1 + x^2} \right)
\sim
\frac{1}{\Gamma} \left( 1 + \frac{y_D^2 - x_D^2}{2} + \cdots \right),
\label{G+}
\\*[3mm]
G_- &\equiv& \int_0^{+ \infty} \!|g_-(t)|^2 dt =
\frac{1}{2 \Gamma} \left( \frac{1}{1 - y^2} - \frac{1}{1 + x^2} \right)
\sim
\frac{1}{\Gamma} \left( \frac{y_D^2 + x_D^2}{2} + \cdots \right),
\label{G-}
\\*[3mm]
G_{+-} &\equiv& \int_0^{+ \infty} \!g_+^\ast(t) g_-(t) dt =
\frac{1}{2 \Gamma} \left( \frac{-y}{1 - y^2} + \frac{-ix}{1 + x^2} \right)
\sim
\frac{1}{\Gamma} \left( \frac{- y_D - i x_D}{2} + \cdots \right).
\label{G+-}
\end{eqnarray}
%

\section{CP-even and CP-odd corrections to the extraction
of $\sin^2 \gamma$}
\label{app:B}

We have seen in Eqs.~(\ref{c+-GWwrong}) and (\ref{c+-ADSwrong}) that
the presence of
mixing may affect $\cos(\gamma \pm \Delta)$ by
\begin{eqnarray}
c_{+w} &=& \cos(\gamma + \Delta) + \delta_e + \delta_o,
\nonumber\\
c_{-w} &=& \cos(\gamma - \Delta) + \delta_e - \delta_o,
\label{c+-w}
\end{eqnarray}
where $\delta_e$ and $\delta_o$ stand for CP-even and CP-odd
corrections, respectively.
The strong phase is given by $\Delta=\Delta_B$ in the GW method,
and by $\Delta = \Delta_B - \Delta_{Di}$ in the ADS method.

For example,
the corrections to the GW method described in Eqs.~(\ref{c+-GWwrong}),
are
\begin{eqnarray}
\delta_e &=&
- \frac{y_D}{2 \tilde{\epsilon}}
	\left[ 2 \tilde{\epsilon}  \eta_f\,
	\cos{(\gamma + 2 \theta_D)} \cos{(\Delta_B)}
	+ \cos{(2 \theta_D)}
\right],
\nonumber\\
\delta_o &=&\
- \frac{x_D}{2 \tilde{\epsilon}} \sin(2 \theta_D)
+ y_D \eta_f\, \sin{(\gamma + 2 \theta_D)} \sin{(\Delta_B)}.
\label{osdeltasGW}
\end{eqnarray}
Notice that the CP-odd term proportional to $x_D$ does not involve
$\Delta$. In fact, that term is due to the interference CP violation
present in the decay $D \rightarrow f_{\rm cp}$,
and it is completely independent of the production mechanism of
the neutral $D$ meson.
Nevertheless,
as we will now show,
when Eqs.~(\ref{c+-w}) are substituted into Eq.~(\ref{trigno-exercise}),
the CP-odd contributions to the
difference between the wrong value of $\sin^2 \gamma_w$
(obtained using $c_{\pm w}$) and the correct value of $\sin^2 \gamma$
(obtained using $\cos(\gamma \pm \Delta)$),
are always proportional to $\sin{\Delta}$.

We start by noting that $c_{\pm w}$ have a standard CP-even (CP-odd)
component given by $\cos{\gamma} \cos{\Delta}$
($- \sin{\gamma} \sin{\Delta}$),
in addition to the mixing component $\delta_e$ ($\delta_o$).
We will denote their sum by
\begin{eqnarray}
c_e &=&
\cos{\gamma} \cos{\Delta} + \delta_e,
\nonumber\\
c_o &=&
- \sin{\gamma} \sin{\Delta} + \delta_o,
\end{eqnarray}
and write
\begin{equation}
c_{+w} = c_e + c_o,\ \ \ \ c_{-w} = c_e - c_o.
\end{equation}
Then
\begin{eqnarray}
\sin^2 \gamma_w
&=&
\frac{1}{2} \left[ 1 - c_{+w} c_{-w} \pm
\sqrt{(1-c_{+w}^2)(1-c_{-w}^2)} \right]
\nonumber\\*[1mm]
&=&
\frac{1}{2} \left[ 1 - c_e^2 + c_o^2  \pm
\sqrt{(1 - c_e^2 + c_o^2)^2 - 4 c_o^2} \right],
\end{eqnarray}
and the results depend only on $c_e^2$ and $c_o^2$.\footnote{Notice
that this property is completely general and holds for any
method in which one is ultimately 
measuring the square of a CP-violating quantity.
Indeed,
the quantity $\sin^2 \gamma_w$ is CP-even (its signs remains the same
under a CP transformation).
Therefore,
CP-even ($c_e$) and CP-odd ($c_o$) contributions to this quantity
cannot interfere with one another and,
moreover,
$c_o$ can only contribute in the combination $c_o^2$.}

Consequently,
the mixing contributions to $c_e^2$ and $c_o^2$ are either
quadratic (and, thus, much suppressed, although they could
dominate for small values of $\cos(\gamma \pm \Delta)$),
or linear,
but appearing only in the combinations $\delta_e \cos{\gamma} \cos{\Delta}$
and $\delta_o \sin{\gamma} \sin{\Delta}$.
Now,
in the GW method $x_D$ only shows up in $\delta_o$,
{\it c.\,f.\,} Eqs.~(\ref{osdeltasGW}).
Therefore,
the mixing contribution to the
GW extraction of $\sin^2 \gamma_w$ proportional to
$x_D \sin(2 \theta_D)$ is also proportional to $\sin \Delta_B$ and,
thus,
it vanishes in the $\Delta_B \rightarrow 0$ limit.

We can also use this discussion to explain why $\sin^2 \gamma_w$
is very similar to $\sin^2 \gamma$ for $\gamma \sim \pi/2$,
provided that $\Delta$ is small.
Indeed,
in that case $c_e^2 = \delta_e^2$,
$c_o^2 = \sin^2{\gamma} \sin^2{\Delta}
	- 2 \delta_o \sin{\gamma} \sin{\Delta}
	+ \delta_o^2$,
and both contributions become small.

The situation is altered in the ADS method because
there we have a new CP-even phase,
$\Delta_{Di}$,
which is expected to be large.
Using Eqs.~(\ref{c+-ADSwrong}) we find the
leading CP-even and CP-odd contributions
to $c_{\pm w}$ to be
\begin{eqnarray}
\delta_e 
&=& 
- \frac{x_D}{2 \epsilon_i} (1 - \epsilon_i^2)
	\cos{(\gamma + 2 \theta_D)} \sin{\Delta_B}
+ \frac{x_D}{2 \tilde{\epsilon}}
	\cos{(2 \theta_D)} \sin{\Delta_{Di}}
\nonumber\\
&&
+ \frac{y_D}{2 \epsilon_i}
	\cos{(\gamma + 2 \theta_D)} \cos{\Delta_B}
- \frac{y_D}{2 \tilde{\epsilon}}
	\cos{(2 \theta_D)} \cos{\Delta_{Di}},
\nonumber\\
\delta_o
&=& 
\frac{x_D}{2 \epsilon_i} (1 - \epsilon_i^2)
	\sin{(\gamma + 2 \theta_D)} \cos{\Delta_B}
- \frac{x_D}{2 \tilde{\epsilon}}
	\sin{(2 \theta_D)} \cos{\Delta_{Di}}
\nonumber\\
&&
- \frac{y_D}{2 \epsilon_i}
	\sin{(\gamma + 2 \theta_D)} \sin{\Delta_B}
+ \frac{y_D}{2 \tilde{\epsilon}}
	\cos{(2 \theta_D)} \cos{\Delta_{Di}},
\end{eqnarray}
respectively.
These are added to the standard contributions
$\cos{\gamma} \cos{(\Delta_B - \Delta_{Di})}$ and
$- \sin{\gamma} \sin{(\Delta_B - \Delta_{Di})}$,
respectively.
The presence of a potentially large $\Delta_{Di}$ has two consequences.
Firstly,
the CP-odd quantity that exists even in the absence of mixing,
is proportional to $\sin{(\Delta_B - \Delta_{Di})}$ and can be large.
Therefore,
the fact that the mixing CP-odd contribution to $\sin^2 \gamma_w$
linear in $\delta_o$ always appears multiplied by 
$\sin{(\Delta_B - \Delta_{Di})}$ ceases to constitute a suppression
factor.
Secondly,
there is now one $x_D$ contribution to $\delta_e$
which is proportional to $\sin{\Delta_{Di}}$.
This will interfere with the standard CP-even contribution,
$\cos{\gamma} \cos{(\Delta_B - \Delta_{Di})}$,
as is also unsuppressed in the $\Delta_B \rightarrow 0$
limit.\footnote{There is also a new CP-even contribution,
$x_D (1 - \epsilon_i^2) \cos{(\gamma + 2 \theta_D)}$,
but it is proportional to $\sin{\Delta_B}$.}

\section{Measuring $\Delta_D$ at a Tau-Charm Factory}
\label{app:C}

\begin{figure}
\centerline{\psfig{figure=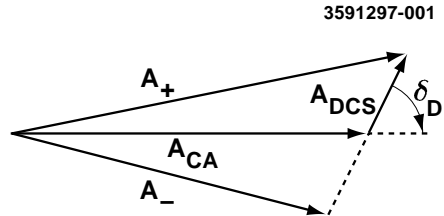,height=1.2in}}
\caption{Obtaining the phase $\Delta_D$ of Equation~(\ref{eq:delta_d})
       from the
	Cabibbo allowed $D^0$~decay amplitude, $A_{CA}$, the doubly Cabibbo
	suppressed amplitude, $A_{DCS}$, and their interference,
	$A_\pm \equiv A_{CA}\pm A_{DCS}$.}
\label{fig:3770-triangle}
\end{figure}

We proceed to study the measurement of $\Delta_D$ at a charm factory,
operating at the $\psi(3770)$ resonance.
For simplicity, we will discuss this measurement in the
context of the SM.
Given the current bounds on $x_D$ and $y_D$,
we expect this measurement to be rather insensitive to
$D^0 - \overline{D^0}$ mixing.
The relation
\begin{equation}
\Delta_D = {\rm arg}\left[A(D^0\rightarrow f_D)
			    A(\overline{D^0} \rightarrow f_D)^*\right]
	\label{eq:delta_d}
\end{equation}
is graphically represented in Figure~\ref{fig:3770-triangle},
demonstrating how to obtain $\Delta_D$ from the the Cabibbo allowed
$D$~decay amplitude, $A_{CA}$,
the doubly Cabibbo suppressed amplitude, $A_{DCS}$,
and their interference, $A_\pm \equiv A_{CA} \pm A_{DCS}$.
While $A_{CA}$ and $A_{DCS}$ have been measured at CLEO~\cite{DtoKpi}
for the $K^-\pi^+$ mode by using $D^{*+}$ decays to tag the 
$D^0$~flavor,
measuring $A_\pm$ requires producing $D^0 - \overline{D^0}$ pairs
in a known coherent state.
It is therefore best to perform all three measurements at the charm factory,
canceling many systematic errors in the construction 
of the triangles of Figure~\ref{fig:3770-triangle}.
To measure the amplitude $A_+$ ($A_-$),
one of the $\psi(3770)$ daughters is tagged as a $D^0_-$
($D^0_+$) by observing it decay into a CP-odd (CP-even) state,
such as $K_S\;\pi^0$ ($K^+K^-$).
The other daughter is then $D^0_+$ ($D^0_-$),
and the fraction of the time that it is seen decaying into $K^-\pi^+$
gives the interference amplitude ${1\over2}|A_\pm|^2$.
We immediately find
\begin{equation}
\cos\Delta_D = \pm {|A_\pm|^2 - |A_{CA}|^2 - |A_{DCS}|^2 \over
	2 |A_{CA}| | A_{DCS}|}.
\end{equation} 
Due to the low statistics tagging scheme of the $|A_\pm|$ measurement
and the fact that $\left|A_{DCS}\right| \ll \left|A_{CA}\right|$,
the error in $\cos\Delta_D$ is dominated by the $|A_\pm|$ measurement
error.
Hence
\begin{eqnarray}
\sigma_{\cos\Delta_D}
	&\approx& {\sigma_{|A_\pm|} \over \left|A_{DCS}\right|}
	\approx {\sigma_{|A_\pm|} \over |A_\pm|} \;
		  \left|{A_{CA} \over A_{DCS}}\right| \nonumber\\
	&=&
		{1 \over 2 \sqrt{N_{A_\pm}}} \;
		  \left|A_{CA} \over A_{DCS}\right|, \label{eq:cos-delta-err2}
\end{eqnarray}
where $N_{A_\pm}$ is the number of events observed in the $A_\pm$
channels,
and we made use of 
$\left|A_{CA}\right| \approx \left|A_\pm\right|$.
Since the event is fully reconstructed,
background is expected to be small,
and its contribution to $\sigma_{\cos\Delta_D}$ is neglected 
in this discussion.
$N_{A_\pm}$ is given by
\begin{eqnarray}
	N_{A_\pm} &=&
	N_{D \overline D}
	\ {\cal B}(D^0\rightarrow K^-\pi^+) \; \epsilon(K^-\pi^+)\nonumber\\
	&\times& \sum_i {\cal B}(D^0 \rightarrow t_i) \; \epsilon(t_i),
	\label{eq:num-ddbar-events}
\end{eqnarray}
where $N_{D\overline D}$ is the number of
$\psi(3770)\rightarrow D^0 \overline D^0$ events,
$t_i$ are the CP-eigenstates used for tagging,
and $\epsilon(X)$ is the reconstruction efficiency of the state $X$.
Taking~\cite{Sof99}
$\sum_i {\cal B}(D^0 \rightarrow t_i) \; \epsilon(t_i) \approx 0.02$,
$\epsilon(K^-\pi^+) \approx 0.8$ and
$\left|A_{CA} / A_{DCS}\right| = 1/\sqrt{0.0031}$, 
and 
${\cal B}(D^0\rightarrow K^-\pi^+) \approx 0.04$,
Eq.~(\ref{eq:cos-delta-err2}) becomes
\begin{equation}
\sigma_{\cos\Delta_D} \approx {370 \over \sqrt{N_{D\overline D}}}.
\label{eq:num_psi3770}
\end{equation}

It is expected that in one year the charm factory will collect
10~$\rm fb^{-1}$,
or $N_{D\overline D} = 2.9\times10^{7}$~\cite{ref:charm-fact},
resulting in $\sigma_{\cos\Delta_D} \approx 0.065$.
Thus,
$\cos\Delta_D$ can be measured to high precision,
even in the presence of background and with relatively modest
luminosity.
We note that the same measurement technique can be used
with multi-body $D^0$~decays,
in which $\cos\Delta_D$ varies over the available phase space.
While the relative statistical error in every small region of
phase space will be large,
its effect on the measurement of $\gamma$ in $B\rightarrow
DK$ will be proportionally small.
The total error in $\gamma$ due to $\Delta \cos\Delta_D$ will be
as small as in the two~body $K^-\pi^+$ mode,
up to differences in $D^0$ branching fractions,
reconstruction efficiencies, and backgrounds.

\end{document}